\documentstyle[12pt,russian]{article}
\newcommand{\diver}{{\rm\, div\, }}
\newcommand{\sign}{{\rm\, sign\, }}
\newcommand{\rot}{{\rm\, rot\, }}

\newcommand{\sun}{{^{\bigodot}}}

\makeatletter

\begin{document}

\input epsf

\begin{center}
\large
{\bf Unipolar Induction of a Magnetized Accretion Disk around a Black Hole}

Published in Astronomy Letters, Vol. 29, No. 3, 2003, pp. 155.159.

A. A. Shatskiy
\footnote{E-mail: shatskiy@lukash.asc.rssi.ru}

Astrospace Center, Lebedev Institute of Physics, Russian Academy of Sciences,
Profsoyuznaya ul. 84/32, Moscow, 117997 Russia
Received October 3, 2002

Abstract

The structure and magnitude of the electromagnetic field 
produced by a rotating accretion disk around a black hole
were determined. The disk matter is assumed to be a 
magnetized plasma with a frozenin
poloidal magnetic field. The vacuum approximation 
is used outside the disk.
\end{center}
$$ $$

Key words: pulsars, neutron stars; black holes, quasars, jets, accretion disks.

\section{INTRODUCTION}

Recently, various models of particle acceleration
near supermassive black holes (SMBHs) in galactic
nuclei and near stellar-mass black holes (BHs) in
the Galaxy have been widely discussed in connection
with the studies of synchrotron radiation and inverse
Compton scattering from narrow-beam jets observed
over a wide spectral range, from radio to gamma rays.
Nevertheless, the particularly high angular resolution
provided by radio interferometers does not allow the
central part of a quasar to be distinguished, suggesting
that the jet width is extremely small (comparable
to the gravitational radius).
Previously (Shatskiy 2001), the mechanism of
Blandford and Znajek (1977) and Blandford (2001)
for electric-.eld generation through the interaction of
the magnetic .eld from a ring current with the gravimagnetic
.eld (GMF) of a Kerr BH located on the
common axis with the ring current was used as the
model of particle acceleration. Bisnovatyi-Kogan and
Blinnikov (1972) and Shatskiy and Kardashev (2002)
considered the mechanism of Deutsch (1955) or
Goldreich and Julian (1969) for electric-field generation
by the unipolar induction produced by the
axial rotation of an accretion disk with a frozen-in
magnetic field. Here, we suggest a mechanism that
combines a unipolar inductor and strong gravitational
SMBHs effects. Naturally, this mechanism
is closer to the actual processes that take place in
quasars.
In contrast to the mechanism from Blandford
and Znajek (1977), Beskin et al. (1992), and Beskin
(1997), the mechanisms of Bisnovatyi-Kogan
and Blinnikov (1972) used previously (Shatskiy 2001;
Shatskiy and Kardashev 2002) operate in the vacuum
approximation; the validity criterion for the latter is
the condition of Goldreich and Julian (1969) for the
number density of free charges:
\begin{equation}
n_e < |({\bf \Omega H})|/(2\pi c e) .
\label{0-1}
\end{equation}
Here, $\Omega$ is the plasma angular velocity, ${\bf H}$ is the
characteristic magnetic field, $c$ is the speed of light,
and $e$ is the elementary charge. Shatskiy and Kardashev
(2002) showed that condition (1) could be
satisfied near a BH, because there are no stable orbits
for the particles closer than three gravitational radii
(in a Schwarzschild field). The matter for which the
vacuum approximation breaks down must be in the
accretion disk, which, because of the effect of Bardeen
and Petterson (1975), must be located in the equatorial
plane of a rotating BH. We do not consider models
in which the vacuum approximation breaks down
in the entire space outside a BH (models with the
magnetohydrodynamic approximation). These were
considered in detail, for example, in the review articles
by Beskin et al. (1992) and Beskin (1997).
Here, we determine the energy of the charged particles
accelerated by the unipolar mechanism as well
as the configurations of the electromagnetic field and
the acceleration region.

\section{CONSTRUCTING THE MODEL}

Consider a Schwarzschild BH surrounded by
an equatorial accretion disk. Let the disk have the
characteristic size $R$ and width $2a$ (see the figure).
Because of the Bardeen-Petterson effect, the disk
thickness can be disregarded.
We use the following notation: $M$ is the mass
of the central body, $R_g = 2M$ is the Schwarzschild radius, 
\footnote{Below, we use the system of units in which the speed of
light and the gravitational constant are equal to unity: c = 1, G = 1.}
$m$ is the mass of the test particle, $u_j$ is its
4-velocity, $F_{ij} = \partial_i A_j - \partial_j A_i$ 
is the electromagnetic-field (EMF) tensor, $A_j$ is the 
EMF potential, and $\Gamma^i_{jk}$
are the Christoffel symbols.
Let us write the Schwarzschild metric and its determinant
in spherical coordinate:
\begin{equation}
\begin{array}{lcr}
ds^2 = (1 - r_g/r) dt^2 - (1 - r_g/r)^{-1} dr^2 - r^2 
d\theta^2 - r^2 \sin^2\theta d\varphi^2\, , \\ 
\\
g = - r^4 
\sin^2\theta \, .
\end{array}
\label{1-0} \end{equation}
In general relativity, the following quantity for a
charged particle that moves in stationary fields is
conserved:
\begin{equation}
\varepsilon = m (u_0 - 1) + e A_0 ,
\label{1-1}
\end{equation}
which matches the particle energy in the nonrelativistic
case. To prove this, it will suffice to consider the
equation of motion for a charged particle in general
relativity (see Landau and Lifshitz 1988):
\begin{equation}
m {du_i\over ds} = m u^k u_l \Gamma^l_{ik} + e u^k F_{ik} ,
\label{1-2}
\end{equation}
where $ds$ is the element of the particle proper time [see
formulas (2)]. After transformation, this expression
reduces to
\begin{equation}
{d\over ds}(m u_i + e A_i) = {m \over 2} u^l u^k 
\partial_i g_{kl} +
e u^k \partial_i A_k .
\label{1-3}
\end{equation}
The conservation of energy . throughout the particle
evolution follows for $i = 0$ in stationary fields.

Since the magnetic field is frozen into the disk,
its distribution inside the disk is determined only by
the initial conditions of the problem. These conditions
depend on the accretion-disk formation mechanism.
If the disk is assumed to have been formed through
the destruction of a star by BH tidal forces, then the
magnetic field of this star in the disk will preserve its
direction. This field can have the profile shown in the
figure.

In the frame of reference comoving with the accretion
disk, there is no electric field inside the disk
because of its conductivity. Therefore, in a fixed frame
of reference (with respect to distant stars), an electric
field is induced by disk rotation inside the disk.
Let the plasma in the disk rotate at an angular
velocity $\Omega$ relative to distant stars. 
The transformation
of coordinates to a rotating frame is then:
\footnote{Unless otherwise specified, 
$x_i = t, r, \theta, \varphi ; x_\alpha = R, \theta$ 
(Greek and Roman indices).}
\begin{equation}
dx^i = d{x'}^k [\delta^i_k + \Omega \delta^i_\varphi 
\delta^0_k ] .
\label{1-5}
\end{equation}
Because of axial symmetry, only the following EMF
potential components are nonzero: $A_0$, the electric-
field potential, and $A_\varphi$, the magnetic-field potential.
According to (6) (see Landau and Lifshitz 1988), the
EMF components transform as
\begin{equation}
A'_0 = A_0 + \Omega A_\varphi , \quad
A'_\varphi = A_\varphi , \quad
F'_{\alpha 0} = F_{\alpha 0} + \Omega F_{\alpha\varphi} , 
\quad
F'_{\alpha\varphi} = F_{\alpha\varphi} .
\label{1-6}
\end{equation}
Since $F_ {\alpha 0} = 0$ in plasma, we have inside the disk
\begin{equation}
F_{\alpha 0} = -\Omega F_{\alpha\varphi} , \quad
A_0 = const - \Omega A_\varphi .
\label{1-7}
\end{equation}
On the disk surface, continuous boundary conditions
exist for the tangential electric-field components and
for the normal magnetic-field components. Outside
the disk, there are no field sources by the definition
of the model. Thus, determining the EMF reduces to
solving the Laplace equation in the spacetime curved
by gravity with the specified boundary conditions on
the disk surface and on the BH horizon. The boundary
conditions for the EMF tensor on the BH horizon
were found previously (Shatskiy 2001):
\begin{equation}
\lim_{r\to r_g} F_{0\theta}\propto g_{00}\to 0 \, , \quad 
\lim_{r\to r_g} F^{r\varphi}\propto g_{00}\to 0 \, .
\label{1-4}
\end{equation}
In turn, the boundary conditions on the disk for the
magnetic and electric fields are determined solely by
the magnetic-field distribution inside it. The specific
form of this distribution is not that important for the
solution of the problem. This is because at distances
from the disk to the point of observation much larger
than the disk thickness, the dipole field (the monopole
field must be absent, because the total disk charge is
zero) mainly contributes to the electric field of the disk
element within position angles between $\varphi$ and 
$\varphi + d\varphi$
when the field is expanded in multipoles. In this case,
the total electric field obtained by integrating over the
angle $\varphi$ has a quadrupole nature:
$$
\lim_{(r/r_g)\to\infty} A_0 = const \cdot (1 - 
3\cos^2\theta )/r^3 .
$$

\section{THE MAXWELL EQUATIONS}

The Maxwell equations for the EMF in general
relativity are
\begin{equation}
{1\over\sqrt{-g}} \partial_i (\sqrt{-g} F^{ik}) = 4\pi 
j^k .
\label{2-1}
\end{equation}
In the disk, the 4-vector of the current $j^k$ can be determined
from a given magnetic field. Let us introduce
the physical components of the EMF vectors, their
analogs in Euclidean space:
\footnote{We denote them by a hat.}
\begin{equation}
{\bf\hat E}^\alpha = - F_{\beta 0} \sqrt{|g^{00} 
g^{\alpha\beta}|} , \quad
{\bf\hat H}_\alpha = 
- e_{\alpha\beta\varphi}F_{\gamma\varphi}
\sqrt{|g^{\gamma\beta}g^{\varphi\varphi}|}
 , \quad
{\bf\hat J}^\alpha = j^\beta \sqrt{|g_{\alpha\beta}|} .
\label{2-2}
\end{equation}
Here, $e^{\alpha\beta\gamma} = e_{\alpha\beta\gamma}$ 
is the Levi-Civita symbol. This
form of the EMF physical components was chosen
in order that Eq. (10) correspond to the classical
Maxwell equations in Euclidean space:
\begin{equation}
\diver {\bf\hat E} =  4\pi j^0 ,\quad
\rot {\bf\hat H} = 4\pi {\bf\hat J} .
\label{2-3}
\end{equation}

\section{THE ELECTROMAGNETIC FIELD NEAR A SMBH}

The magnetic field of an accretion disk around
a Schwarzschild BH was determined by Tomimatsu
and Takahashi (2001). The electric field of the disk
element within position angles between 
$\varphi$ and $\varphi + d\varphi$
can be represented as the field from two charges:
$+q{d\varphi\over 2\pi}$
and 
$-q{d\varphi\over 2\pi}$,
located inside the disk, at the
system equator, and at distances $+a$ and $-a$ from
its center ($r = R, \theta. = \pi /2$), respectively
\footnote{Naturally, there are no free charges in the disk; these were
introduced for the convenience of representing the dipole field
outside the disk.}.
As a result,
we obtain an electric dipole $2qad\varphi /\pi$ in the
disk element between $\varphi$ and $\varphi + d\varphi$. 
In Euclidean
space, the radial electric field at the disk center can
be obtained by integrating over the angle $\varphi$; 
for $a << R$, it is $E_0 = \hat E^r_{(r = R, \theta = \pi /2)} 
= -2q/(\pi Ra)$. The corresponding
magnetic field (which is responsible for the
emergence of the electric field) can be found from the
electric field. Note that nothing forbids the frozenin
magnetic field in the disk to have precisely such a
profile (see the figure).
In the figure, the accretion disk is located at 
$r \approx 6M$. The contradictions related to the existence of
stable orbits in this region can be removed by the
following reasoning:

1  For the Kerr metrics, the nearest stable orbit is
at radius $r = M$ (see, e.g., Landau and Lifshitz 1988).

2  Even if the orbit is not stable, it is spiral and
goes under the horizon, while a new orbit can arrive
in place of it. Thus, the pattern is quasi-stationary.
According to (11) and (8), the quantity $q$ can be
expressed in terms of the magnetic field at the disk
center 
\footnote{$H_0 = -\hat H_\theta {}_{(r = R, 
\theta = \pi /2)}/\sqrt{1 - r_g/R}$. }
as follows:
\begin{equation}
q = -\pi R a E_0 /2 = {\pi\over 2} \Omega R^2 a H_0 .
\label{3-1}
\end{equation}
In physical coordinates, the field of a point charge e
near a BH was presented by Thorne et al. (1998). It
was obtained in a closed form by Linet (1976):
\begin{equation}
\begin{array}{lll}
A_0 = {e\over Rr}\left[ M + {(R - M)(r - M) - M^2 t \over 
D }
\right] 
\, , \\ 
\\
\hat E^r = {e\over Rr^2} \left\{ M \left[
1 - {R - M + M t \over D } \right] + 
{r \left[ (r - M)(R - M) - M^2 t \right]
\left[ r - M - (R - M) t \right] \over D^3 } \right\}
\, , \\
\\
\hat E^\theta = - {e (R - 2M) \sqrt{1 - 2M/r} \over D^3 } 
\partial_\theta t
\, , \\
\\
D^2 = (r - M)^2 + (R - M)^2 - M^2 - 
2(r - M)(R - M) t + M^2 t^2 
\, .
\end{array}
\label{3-2}
\end{equation}
where t is the cosine of the angle between the directions
of the point charge and the point of observation
of the field from the BH center.
The electric field of a charged ring at the BH equator
was found by Bicak and Dvorak (1996) in the form
of a series. Here, this field is found in a quadrature
form. To this end, we make the following substitutions:
$t \to \sin\theta \cos\varphi$, $e \to Q{d\varphi\over 2\pi}$
and integrate over
the angle from $-\pi$ to $+\pi$. The model electric-field
potential $A^{tot}_0$
is a superposition of the fields from
two charged rings at radii $R + a$ and $R - a$ and with
charges $+q$ and $-q$, respectively. The quadrature obtained
can be expressed in terms of incomplete elliptic
integrals. Since this quadrature is cumbersome, it
makes no sense to write it here. Instead, we give an
expression more useful for practical calculations, an
expansion of this quadrature in terms of $a$. We retain
only the first term of the series (because $a << R$ is small):
\footnote{Here, we make use of the symmetry of the potential $A_0$
in variables $R$ and $r$ and use their change (the subscript
$"{}_{(R \leftrightarrow r)}"$).}
$$
A^{tot}_0  = \int\limits_{-\pi}^{+\pi}
2a\partial_R A_0 d\varphi =
\int\limits_{0}^{+\pi} 4a (\partial_r A_0)_{ (R \leftrightarrow r)}\,
d\varphi 
= \int\limits_{0}^{+\pi}4a \hat E^r_{(R \leftrightarrow r)}\, d\varphi .
$$
Substituting expression (14) here finally yields
\begin{equation}
\begin{array}{lcr}
A_0^{tot} = {2 a q\over \pi r R^2}\int\limits_0^\pi d\varphi
\left( {M\left (D - r + M - Mt\right)\over D} +
{R\left[ (r - M)(R - M) - M^2t\right]\cdot
\left[ R - M - (r - M)t\right] \over D^3} \right) .
\end{array}
\label{3-2-1}\end{equation}
On the $\Omega$ axis, the integration over $\varphi$ is simple and it
is easy to see that the only maximum of the potential
$A^{tot}_0$
(outside the BH) is on the horizon $(r = 2M)$:
\begin{equation}
|A_0^{tot}(r=2M, \theta = 0)|_{_{MAX}} = 
{\pi\Omega a^2 R^2 H_0 \over R^2 - a^2}.
\label{3-3}
\end{equation}

%
\begin{figure}[t]
\centering
\epsfbox[30 180 500 630]{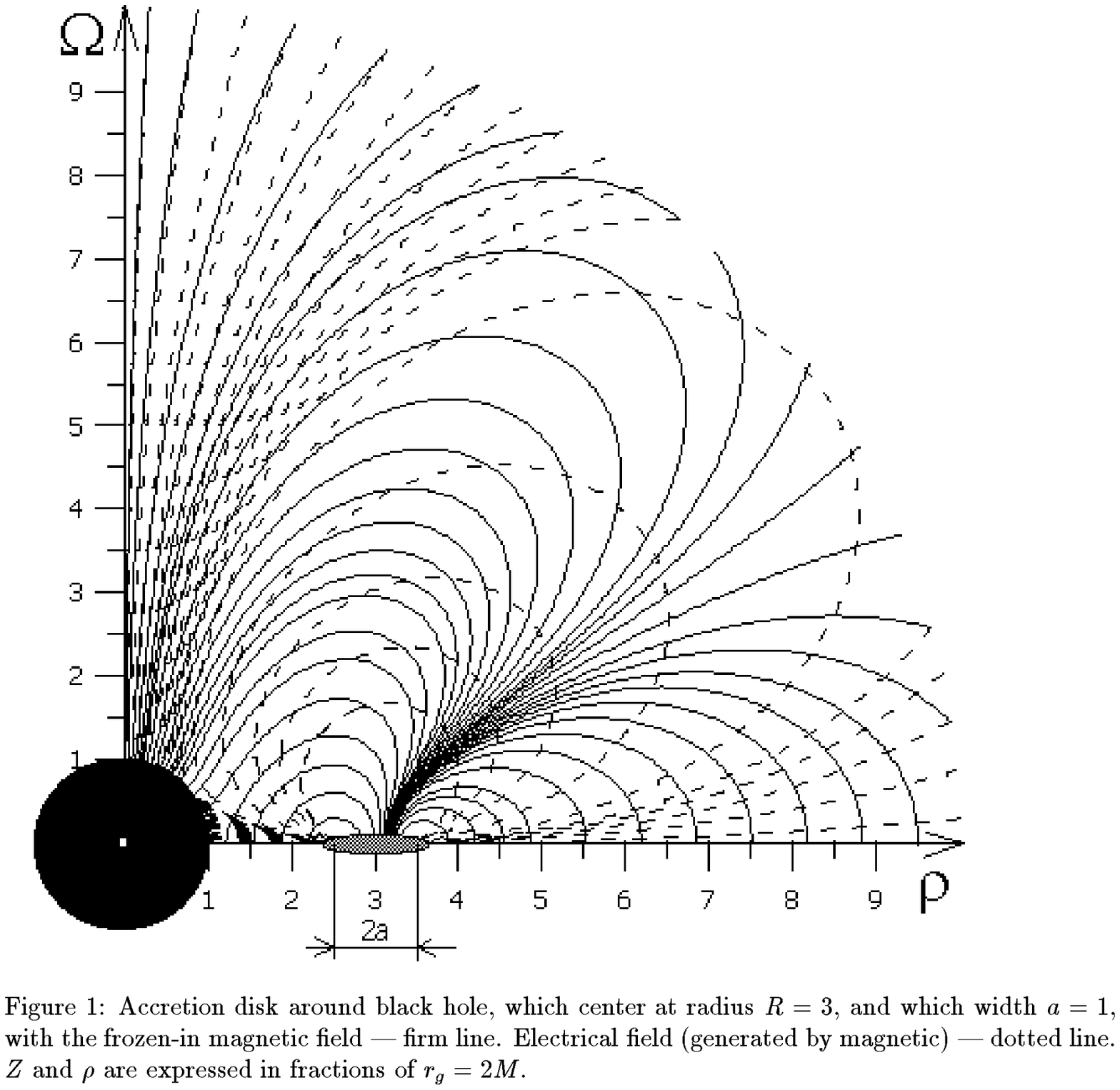}
\label{r1}
\end{figure}
%
Consider the distribution of energies $\varepsilon (r)$ in an ensemble
of test particles at rest on the $\Omega$ axis [see (3)]. 
This distribution has a weak maximum near the
horizon and then slowly falls o. to zero at distances
$r >> r_g$. Subsequently, the energy variation can be
virtually disregarded.

\section{DISCUSSION}

The necessary condition for particle escape to 
infinity from rest is positiveness of the force 
$m{du^r\over ds}= m{d^2r\over ds^2}$ 
that acts on the test particle at the starting
point. This requires that the charge of the test particle
$e$ have the sign opposite to that of $q$:
\begin{equation}
-\sign (eq)=\sign (e{\bf \Omega \hat E_{(\theta =0)}}) = 
-\sign (e{\bf \Omega H_0}) = 1.
\label{3-3-1}\end{equation}
The force that acts on an particle is proportional to the
particle energy gradient (see Eq. (4)]:
\begin{equation}
m{du^r \over ds} = -(\partial_r \varepsilon ) \cdot 
\sqrt{1 - 2M/r} .
\label{3-4}	
\end{equation}

In the main order in $\alpha = mM/(eq) << 1$, the energy
maximum is almost equal to the electric component
of the particle energy on the horizon:
\footnote{At $a \sim r_g \sim R/3$ [see (16)].}
\begin{equation}
\varepsilon_{max} = \left( {1\over\alpha} \cdot {2M 
a\over R^2 - a^2} 
- 1\right) \cdot m c^2 \sim 0.1 \cdot m c^2 /\alpha 
\label{3-5}
\end{equation}
and the point at which this maximum is reached is at
the following distance from the horizon:
\begin{equation}
r_{max} - r_g = r_g \cdot \alpha^2 \cdot {(R^2 - a^2)^2 
[(R - M)^2 - a^2]^4
\over 16 M^6 a^2 (3R^2 + M^2 - 4MR + a^2)^2} 
\sim r_g \cdot 10^3 \cdot \alpha^2 .
\label{3-6}
\end{equation}
In conclusion, several more words can be said
about the same model in a Kerr field. BH rotation
gives rise to a GMF that interacts with the EMF of
the disk and changes its components in magnitude
and direction. In the linear approximation in BH
angular velocity, the GMF gives an additive component
to expression (19) for the maximum energy
of the charged particle accelerated by an electric
field. This component was determined previously
(Shatskiy 2001; it is convenient to represent it here as
\begin{equation}
\varepsilon_{g} \approx {1\over 2\pi^2 } \cdot
\left( {\Omega_g r_g\over \Omega R}\right) \cdot
\left( {r_g^2\over aR}\right) \cdot m c^2 / \alpha \sim 
\varepsilon_{max}.
\label{3-8}\end{equation}
Here, $\Omega_g$ is the angular velocity of the BH horizon 
(the falling test particles are drawn into rotation by the BH
GMF).We see from (19) that at
\begin{equation}
\alpha^{-1} \approx \left({\Omega R\over c}\right)\cdot
\left({aR\over r_g^2}\right)\cdot \left({H_0\over 10^4 
Gauss}\right) \cdot
\left({M\over 10^9 M_\sun}\right) \cdot \left({m_e\over 
m}\right)
\cdot 10^{15} 
\label{3-7}
\end{equation}
the particle energy 
\footnote{Here, $m_e$ is the electron mass.}
accelerated by SMBHs can reach
values larger than $10^{20}$eV.

\section{CONCLUSIONS}

1  The model described above yields even a higher
energy of the accelerated particles than does the
Blandford-Znajek model.

2  The energy excess can be expended on the
radiation reaction and on collisions with particles of
the rarefied plasma near the axis.

3  The derived EMF configuration can be used
to numerically calculate the dynamics of the charged
particles far from the axis. It will make it possible to
compare theoretical conclusions with observational
data.
(4) Far from the axis, a force-free field with the
approximation of magnetohydrodynamic models can
give a large contribution to the EMF amplitude.
Therefore, in numerical calculations, the contributions
to the EMF configuration from different models
should be taken into account with different weights.
$$ $$
\hrule
$ $

This work was supported by the Russian Foundation
for Basic Research 
(project nos. 01-02-16812, 00-15-96698, 01-02-17829). 

\section{ACKNOWLEDGMENTS}

It wish to thank
N.S. Kardashev, R.F. Polishchuk, V.N. Lukash,
B.V. Komberg, Yu.Yu. Kovalev, the remaining sta. of
the theoretical departments at the Astrospace Center
and the Lebedev Physical Institute, and participants
of workshops for an active participation in preparing
the paper and for important remarks.

\section{REFERENCES}
.

[1]. J.M.Bardeen and J. A. Petterson,Astrophys. J. Lett.
195, L65, 1975.

[2]. V. S. Beskin, Usp. Fiz. Nauk 167, 689, 1997.

[3]. V. S. Beskin, Ya. N. Istomin, and V. I. Par'ev, Astron.
Zh. 69, 1258, 1992.

[4]. G. S. Bisnovatyi-Kogan and S. I. Blinnikov, Astrophys.
Space Sci. 19, 119, 1972.

[5]. J. Bicak and L. Dvorak, General Relativity and Gravitation
7, 959, 1996.

[6]. R. Blandford and R. Znajek, MNRAS 179, 433, 1977.

[7]. R. Blandford, Galaxies and Their Constituents at the
Highest Angular Resolution. Proc. IAI Symp. 205.
Ed. R. T. Schilizzi (San Francisco, ASP, 2001), p. 10.

[8]. P. Goldreich andW. H. Julian, Astrophys. J. 157, 869, 1969.

[9]. J. Deutsch, Ann. D'Astrophys. l, 1, 1955.

[10]. L. D. Landau and E.M. Lifshitz, Field Theory (Nauka,
Moscow, 1988).

[11]. B. Linet, J. Phys. A 9, 1081, 1976.

[12]. A. Tomimatsu and M. Takahashi, Astrophys. J. 552,
710, 2001.

[13]. K. Thorne, P. Price, and D. MacDonald (Eds.),
Black Holes. A Membrane Approach (Mir,Moscow, 1998).

[14]. A. A. Shatskiy, Zh. Eksp. Teor. Fiz. 93, 920, 2001, (gr-qc/0202068).

[15]. A. A. Shatskiy and N. S. Kardashev, Astron. Zh. 46,
639, 2002, (astro-ph/0209465).

$ $

\hrule

$ $

Translated by G. Rudnitskii

\end{document}